\documentclass{article}
\usepackage[affil-it]{authblk}
\usepackage{geometry}
\usepackage{url}
\usepackage{amsmath,amssymb,amsfonts,graphicx,amsthm,nicefrac,mathtools,bbm,bm}
\usepackage{hyperref}
\usepackage{verbatim}
\usepackage[numbers]{natbib}
\usepackage{algorithm}
\usepackage{algorithmicx}
\usepackage{algpseudocode}
\usepackage{booktabs,multirow}
\usepackage{fixltx2e}

\title{Learning Nonparametric Forest Graphical Models with Prior Information}

\author{Yuancheng Zhu, Zhe Liu}
\affil{University of Chicago}
\author{Siqi Sun}
\affil{Toyota Technological Institute at Chicago}
\date{}

\newcommand\numberthis{\addtocounter{equation}{1}\tag{\theequation}}
\renewcommand\hat\widehat

\DeclareMathOperator*{\argmax}{arg\,max}
\newcommand\map{\text{{\sc map}}}
\algdef{SE}[DOWHILE]{Do}{doWhile}{\algorithmicdo}[1]{\algorithmicwhile\ #1}%

\begin{document}
\maketitle

\begin{abstract}
We present a framework for incorporating prior information into nonparametric estimation of graphical models. 
To avoid distributional assumptions, we restrict the graph to be a forest and build on the work of forest density estimation (FDE).
We reformulate the FDE approach from a Bayesian perspective, and introduce prior distributions on the graphs.
As two concrete examples, we apply this framework to estimating scale-free graphs 
and learning multiple graphs with similar structures. 
The resulting algorithms are equivalent to finding a maximum spanning tree of a weighted graph with a penalty term on the  connectivity pattern of the graph.
We solve the optimization problem via a minorize-maximization procedure with Kruskal's algorithm. 
Simulations show that the proposed methods outperform competing parametric methods, 
and are robust to the true data distribution. 
They also lead to improvement in predictive power and interpretability in two real data sets. 
\end{abstract}

\section{Introduction}
Graphical models are widely used to encode the conditional independence relationships between random variables. In particular, a random vector $X=(X_1,\dots,X_d)$ is represented by an undirected graph $G=(V,E)$ with $d=|V|$ vertices and missing edges $(i,j)\not\in E$ whenever $X_i$ and $X_j$ are conditionally independent given the other variables. One major statistical task is to learn the graph from $n$ i.i.d.\ copies of the random vector. 

Existing approaches for estimating graphical models make assumptions on either the underlying distribution or the graphical structure. 
Currently the most popular method, called \emph{graphical lasso} \cite{friedman2008sparse}, assumes that the random vector follows a multivariate Gaussian distribution. 
In this way, learning the graph is equivalent to estimating the precision matrix $\Omega$, since the conditional independence of a Gaussian random vector is entirely determined by the sparsity pattern of $\Omega$. 
The graphical lasso finds a sparse estimate of $\Omega$ by maximizing the $\ell_1$-regularized log-likelihood.
On the other hand, we can make no distributional assumptions but restrict the graph to be a forest instead.
Under this structural constraint, there exists a factorization of the density function involving only the univariate and bivariate marginal densities, which makes nonparametric estimation tractable in high dimensions. 
In this case, estimating the graph amounts to finding the maximum spanning tree of a weighted graph; see, for example, \cite{chow1968approximating,liu2011forest} for details. 

Oftentimes, additional information on the structure of a graph is available \emph{a priori}, 
which could be utilized to assist the estimation task. 
For example, a wide variety of the networks in recent literature, such as protein, gene, and social networks, are reported to be scale-free. 
That is, the degree distribution of the vertices follows a power law: $p(\text{degree}=k)\propto k^{-\alpha}$ for some $\alpha>1$.
In such scale-free networks, some vertices have many more connections than others, and these highest-degree vertices are usually called hubs and serve significant roles in their networks. 
As another example of prior information, consider the applications where we believe that several networks share similar but not necessarily identical structures.
This phenomenon is not unusual when we have multiple sets of data across distinct classes or units, such as gene expression measurements collected on a set of normal tissue samples and a set of cancer tissue samples. 
It is thus natural to ask whether such prior information can be integrated to improve the estimation.

Various approaches have been proposed to incorporate the prior belief of the underlying graphs,
for example,~\cite{defazio2012convex,liu2011learning,tan2014learning,tang2015learning} for learning scale-free graphical models, and 
\cite{Guo11,Danaher14,Peterson15,Zhu15} for joint estimation of multiple graphical models.
Nevertheless, to the best of our knowledge, all the existing methods assume some parametric distribution on the data, mostly multivariate Gaussian.
Such distributional assumptions can be quite unrealistic and unnecessary in many applications. 
Even though the marginal distribution of each variable can be transformed to approximately Gaussian, which allows arbitrary univariate distributions, the joint dependence is still restricted under the Gaussian assumption. 

In this paper, we relax such distributional assumptions and estimate graphical models nonparametrically. 
We build on the forest density estimation (FDE) method introduced in \cite{liu2011forest}.
In particular, we reformulate the FDE approach from a Bayesian perspective, 
and encode the prior information by putting some prior distribution on the graphs, 
which favors those that are more consistent with our prior belief.
We further show that for the scale-free-graph case and the multiple-graph case,
such an approach amounts to finding a maximum spanning tree of a weighted graph with a penalty term
on the connection pattern of the nodes. 
We then devise an algorithm based on a minorize-maximization procedure and Kruskal's algorithm \cite{kruskal1956shortest} to find a local optimal solution. 

The rest of the paper is organized as follows. 
In the following section, we give background on forest density estimation.
In Section~3, we first give a general framework on how to incorporate prior information
to nonparametric forest-based graphical model estimation, and then 
illustrate how the framework can be specialized to model scale-free graphical models and
jointly estimate multiple graphical models with similar structure. 
In Section~4, we provide a brief review on the related work.
Experimental results on synthetic data sets and real applications are presented in Section~5, 
followed by a conclusion in Section~6.

\section{Forest density estimation}\label{sec:fde}

We say an undirected graph is a forest if it is acyclic. Let $F=(V_F,E_F)$ be a forest with vertices $V_F=\{1,\dots,d\}$ and edge set $E_F\in V_F\times V_F$. Let $X=(X_1,\dots,X_d)$ be a $d$-dimensional random vector with density $p(x)>0$. We say that $X$, or equivalently, its density $p$, is Markov to $F$ if $X_i$ and $X_j$ are conditionally independent given the other random variables whenever edge $(i,j)$ is missing in $E_F$. A density $p$ that is Markov to $F$ has the following factorization 
\begin{equation}
p(x)=\prod_{(i,j)\in E_F}\frac{p_{ij}(x_i,x_j)}{p_i(x_i)p_j(x_j)}\prod_{\ell\in V_F}p_\ell(x_\ell),\label{factorization}
\end{equation}
where each $p_{ij}(x_i,x_j)$ is a bivariate density and each $p_l(x_l)$ is a univariate density. 
With this factorization, we can write the expected log-likelihood as
\begin{align}
\mathbb E\log p(X)
&= \int p(x)\left(\sum_{(i,j)\in E_F}\log\frac{p_{ij}(x_i,x_j)}{p_i{x_i}p_j(x_j)}+\sum_{\ell\in V_F}\log p_\ell(x_\ell)\right)dx \\
&=\sum_{(i,j)\in E_F}I(X_i;X_j)-\sum_{\ell\in V_F}H(X_\ell),\label{Elogp}
\end{align}
where 
$I(X_i;X_j)=\int p_{ij}(x_i,x_j)\log\frac{p_{ij}(x_i,x_j)}{p_i(x_i)p_j(x_j)}dx_idx_j$ 
is the mutual information between $X_i$ and $X_j$, 
and $H(X_\ell)=-\int p_\ell(x_\ell)\log p_\ell(x_\ell)dx_\ell$ is the entropy of $X_\ell$.
We maximize the right hand side of \eqref{Elogp} to find the optimal forest $F$
\begin{equation}
\hat F=\argmax_{F\in\mathcal F_d} \sum_{(i,j)\in E_F}I(X_i;X_j),\label{Fstar}
\end{equation}
where $\mathcal F_d$ is the collection of spanning trees on vertices $\{1,\dots,d\}$. 
We let $\mathcal F_d$ contain only spanning trees because there is always a spanning tree that solves the problem \eqref{Fstar}.
This problem can be recast as the problem of finding a maximum spanning tree for a weighted graph, where the weight $w_{ij}$ of the edge between nodes $i$ and $j$ is $I(X_i;X_j)$. Kruskal's algorithm \cite{kruskal1956shortest} is a greedy algorithm that is guaranteed to find an optimal solution, while \citet{chow1968approximating} propose the procedure in the setting of discrete random variables. The method is described in Algorithm \ref{alg:kruskal}.

\begin{algorithm}[H]
\begin{algorithmic}
\State \textbf{Input} Weight matrix $W = (w_{ij})_{d\times d}$
\State Initialize $E^{(0)}\gets\emptyset$
\For {$\ell=1,\dots,d-1$}
    \State $(i^{(\ell)},j^{(\ell)})\leftarrow\argmax_{(i,j)}w_{ij}$ such that $E^{(\ell-1)}\cup \{(i^{(\ell)},j^{(\ell)})\}$ doesn't contain a cycle
    \State $E^{(\ell)}\leftarrow E^{(\ell-1)}\cup \{(i^{(\ell)},j^{(\ell)})\}$ 
\EndFor
\State \textbf{Output} The final edge set $E^{(d-1)}$
\end{algorithmic}
\caption{Kruskal's (Chow-Liu) algorithm}
\label{alg:kruskal}
\end{algorithm}

However, this procedure is not practical since the true density $p$ is unknown.
Suppose instead that we have $X_{1,1:d},\dots, X_{n,1:d}$, which are $n$ i.i.d.\ copies of the random vector $X$. We replace the population mutual information by the estimates 
\begin{equation}
\hat I(X_i;X_j)= \int \hat p_{ij}(x_i,x_j)\log\frac{\hat p_{ij}(x_i,x_j)}{\hat p_i(x_i)\hat p_j(x_j)}dx_idx_j,\label{mutualinfo}
\end{equation}
where $\hat p_{ij}(x_i,x_j)$ and $\hat p_\ell(x_\ell)$ are kernel density estimators of the bivariate and univariate marginal densities
\begin{align}
\hat p_{ij}(x_i,x_j)=\frac{1}{n}\sum_{t=1}^n\frac{1}{h_2^2}K\left(\frac{X_{ti}-x_i}{h_2}\right)K\left(\frac{X_{t j}-x_j}{h_2}\right), \ 
\hat p_{\ell}(x_\ell)=\frac{1}{n}\sum_{t=1}^n\frac{1}{h_1}K\left(\frac{X_{t\ell}-x_\ell}{h_1}\right)
\end{align}
with a kernel function $K$ and bandwidths $h_2$ and $h_1$. 
The resulting estimator of the graph becomes
\begin{equation}
\tilde F=\argmax_{F\in\mathcal F_d} \sum_{(i,j)\in E_F}\hat I(X_i;X_j).\label{Fstarhat}
\end{equation}
A held-out set is usually used to prune the spanning tree $\tilde F$ by stopping early in Algorithm \ref{alg:kruskal} when the likelihood on the held-out set is maximized. Thus we obtain a forest estimate of the graph.

\section{Learning forest graphical model with prior knowledge}

\subsection{A Bayesian framework}\label{sec:bayesian-framework}

Sometimes we have some prior information about the structure of the underlying graphical models,
and would like to incorporate that to assist the estimation.
One way to realize that is to encode the prior knowledge into prior distributions on the spanning trees. 
Let $\pi(F)$ be a prior distribution on $\mathcal F_d$, the set of the spanning trees with $d$ nodes.
Given the data $X_{1,1:d},\dots, X_{n,1:d}$ and assuming the density $p$ is known and Markov  to the spanning tree $F$, we can write the likelihood as
\begin{align}
p(X|F)=\prod_{t=1}^n\left(\prod_{(i,j)\in E_{F}}\frac{p_{ij}(X_{ti},X_{tj})}{p_i(X_{ti})p_j(X_{tj})}\prod_{\ell\in V_{F}}p_\ell(X_{t\ell})\right).
\end{align}
Then the posterior probability of $F$ is
\begin{align}
p(F|X)\propto p(X|F)\pi(F)\propto \prod_{t=1}^n\left(\prod_{(i,j)\in E_{F}}\frac{p_{ij}(X_{ti},X_{tj})}{p_i(X_{ti})p_j(X_{tj})}\prod_{k\in V_{F}}p_k(X_{tk})\right)\cdot\pi(F).\label{eqn:posterior}
\end{align}
The maximum a posteriori (MAP) estimate is given by 
\begin{equation}
\hat F_{\map} =\argmax_{F\in\mathcal F_d}\left\{\sum_{(i,j)\in E_F}\sum_{t=1}^n\frac{1}{n}\log \frac{p_{ij}(X_{ti},X_{tj})}{p_i(X_{ti})p_j(X_{tj})}+\frac{1}{n}\log\pi(F)\right\}.\label{Fmap}
\end{equation}
Since we do not know the true density $p$ in practice, we can plug in the estimator \eqref{mutualinfo} and obtain
\begin{equation}
\tilde F_\pi=\argmax_{F\in\mathcal F_d}\left\{\sum_{(i,j)\in E_F}\hat I(X_i;X_j)+\frac{1}{n}\log\pi(F)\right\} \label{Flambdahat}
\end{equation}
as an approximation of $\hat F_\map$.
In fact, $\tilde F_\pi$ is obtained by replacing the true marginal densities and the empirical distributions in \eqref{Fmap} by their corresponding density estimates.
It can also be viewed as a penalized version of the estimator \eqref{Fstarhat}.

The penalty term $\frac{1}{n}\log \pi(F)$, which is sometimes combinatorial, could make the optimization problem extremely hard to solve. 
However, when $\log \pi(F)$ is convex with respect to the entries of the adjacency matrix of $F$, 
we can adopt a minorize-maximization algorithm \cite{hunter2004tutorial} to find a local optimal solution. 
In fact, given the convexity of $\log \pi(F)$, the objective function adopts a linear lower bound at any current estimates.
This linear lower bound can be then decomposed into a sum of weights over the edges, and
we can apply the Kruskal's algorithm to update our estimate. 
We shall see in details in the following two concrete examples how this can be carried out. 

\subsection{Scale-free graphs}
Now suppose that we have reasons to believe that the graph is scale-free, or more generally, 
that the graph consists of several nodes that have dominating degrees compared to the rest. 
Let $\delta(F,l)$ be the degree of the node $l$ of a spanning tree $F\in\mathcal F_d$. Consider a prior distribution on $F$ which satisfies
\begin{equation}
\pi(F)\propto \prod_{\ell\in V_F}\delta(F,\ell)^{-\alpha},
\end{equation}
for some $\alpha>1$.
This prior distribution favors the spanning trees whose degrees have a power law distribution, and thus reflects our prior beliefs. 
Plugging this in \eqref{Flambdahat}, we obtain
\begin{equation}
\tilde F_\pi=\argmax_{F\in\mathcal F_d}\left\{\sum_{(i,j)\in E_F}\hat I(X_i;X_j)-\lambda\sum_{\ell\in V_F}\log(\delta(F,\ell))\right\}, \label{eqn:Ftilde-scalefree}
\end{equation}
where $\lambda = \alpha/n$ can be now viewed as a tuning parameter.
To solve this optimization problem, we first rewrite the objective function as
\begin{align}
f(F)= \sum_{i<j}w_{ij}F_{ij}-\lambda\sum_{i=1}^d\log\left(\sum_{j=1}^d F_{ij}\right),
\end{align}
where $w_{ij}=\hat I(X_i;X_j)$. 
Here we also abuse our notation by writing $F$ as the adjacency matrix of $F$,
that is, $F_{ij}=1$ if and only if $(i,j)\in E_F$.
Note that we have the additional constraint that the graph $F$ is a spanning tree. 
Given a current estimate $\check F$,
we first lower bound $f(F)$ by linearizing it at $\check F$:
\begin{align}
f(F)
&\geq \sum_{i<j}w_{ij}F_{ij}- \lambda\sum_{i=1}^d\left(\log\left(\sum_{j=1}^d \check F_{ij}\right) +\frac{\sum_{j=1}^dF_{ij}-\sum_{j=1}^d\check F_{ij}}{\sum_{j=1}^d\check F_{ij}}\right)\\
&=\sum_{i<j}\left(w_{ij}-\frac{\lambda}{\sum_{\ell=1}^d\check F_{i\ell}}-\frac{\lambda}{\sum_{\ell=1}^d\check F_{j\ell}}\right)F_{ij}+C,\numberthis\label{lowerbound}
\end{align}
where $C$ is a constant which doesn't depend on $F$. 
We can maximize this lower bound by applying Kruskal's algorithm to the graph with edge weights
\begin{equation}
\check w_{ij}=w_{ij}-\frac{\lambda}{\sum_{\ell=1}^d\check F_{i\ell}}-\frac{\lambda}{\sum_{\ell=1}^d\check F_{j\ell}}.
\label{weightupdate}
\end{equation}
We see that the weights are updated at each iteration based on the current estimate of the graph. 
Each edge weight is penalized by two quantities that are inversely proportional to the degrees of the two endpoints of the edge. 
An edge weight is thus penalized less if its endpoints are already highly connected and vice versa. 
With such a ``rich gets richer'' procedure, the algorithm encourages some vertices to have high connectivity and hence the overall degree distribution to have a heavy tail. 
We iterate through such minorization and maximization steps until convergence.
Since the objective function is always increasing, the algorithm is guaranteed to converge to a local maximum. 

\subsection{Multiple graphs with similar structure}
In this part, we illustrate how the framework can be modified to 
facilitate the case where we have multiple graphs that are believed
to have similar but not necessarily identical structures. 
Instead of one single graph, suppose that we now have $K$ graphical models 
with underlying forests $F^{(1)},\dots,F^{(K)}$,
and for the $k$th one, we observe data $X^{(k)}=(X_{1,1:d},\dots,X_{n_k,1:d})$.
Given a joint prior distribution $\pi$ on $(F^{(1)},\dots,F^{(K)})$, 
we combine the likelihood for the $K$ models and update
the posterior distribution \eqref{eqn:posterior} to be
\begin{equation}\label{eqn:posterior-multiple}
\begin{aligned}
p(F^{(1:K)}|X^{(1:K)})\propto\prod_{k=1}^K\prod_{t=1}^n\left(\prod_{(i,j)\in E_{F^{(k)}}}\frac{p^{(k)}_{ij}(X_{ti}^{(k)},X_{tj}^{(k)})}{p_i^{(k)}(X_{ti}^{(k)})p_j^{(k)}(X_{tj}^{(k)})}\prod_{\ell\in V_{F^{(k)}}}p_\ell^{(k)}(X_{t\ell}^{(k)})\right)\cdot\pi(F^{(1:K)}).
\end{aligned}
\end{equation}
Next, we design a prior distribution on the set of $K$ spanning trees which reflects
our belief that the structures across the $K$ of them share some similarity.
Again we use $F^{(k)}$ to denote the adjacency matrix
of the corresponding graph, that is, $F^{(k)}_{ij}=1$ if and only if $(i,j)\in E_{F^{(k)}}$.
We consider the following hierarchical model:
\begin{align}
\tau_{ij}&\sim \text{Beta}(\alpha,\beta) \text{ for all }i<j,\\
F_{ij}^{(k)}\,|\,\tau_{ij}&\sim\text{Bernoulli}(\tau_{ij}) \text{ for all $k$ and $i<j$}.
\end{align}
According to this model, the same edge across multiple graphs is governed by the same
parameter $\tau_{ij}$, and hence encourage similarity across them.
This essentially gives a prior distribution on $F^{(1:K)}$:
\begin{align}
\pi(F^{(1:K)})&\propto\prod_{i<j}\int_{\tau_{ij}}p(F_{ij}\,|\,\tau_{ij})p(\tau_{ij})d\tau_{ij}\cdot\mathbbm{1}\{F^{(k)}\in\mathcal{F}_d\text{ for all }k\} \\
&=\prod_{i<j}\int_{\tau_{ij}}\left[\prod_{k=1}^K p(F_{ij}^{(k)}|\tau_{ij})\right]p(\tau_{ij})d\tau_{ij}\cdot\mathbbm{1}\{F^{(k)}\in\mathcal{F}_d\text{ for all }k\} \\
&\propto\prod_{i<j}\int_{\tau_{ij}}\tau_{ij}^{\alpha+\|F_{ij}\|_1-1}(1-\tau_{ij})^{\beta+K-\|F_{ij}\|_1-1}d\tau_{ij}\cdot\mathbbm{1}\{F^{(k)}\in\mathcal{F}_d\text{ for all }k\} \\
&= \prod_{i<j}B(\alpha+\|F_{ij}\|_1,\beta+K-\|F_{ij}\|_1)\cdot\mathbbm{1}\{F^{(k)}\in\mathcal{F}_d\text{ for all }k\},
\end{align}
where $F_{ij}$ is the vector containing the $(i,j)$th entries of $F^{(k)}$ for $k=1,\dots,K$,
$\|\cdot\|_1$ denotes the $\ell_1$ norm, and $B(\cdot,\cdot)$ denotes the Beta function.
Now combining this with \eqref{eqn:posterior-multiple} 
and following the reasoning in Subsection \ref{sec:bayesian-framework}, 
we obtain our estimator in this case
\begin{equation}\label{eqn:Ftilde-multiple}
\begin{aligned}
\tilde F^{(1:K)}_\pi=\argmax_{F^{(k)}\in\mathcal F_d,\, \forall k}\left\{\sum_{k=1}^K\sum_{(i,j)\in E_{F^{(k)}}}\hat I(X_i^{(k)};X_j^{(k)})
+\lambda\sum_{i<j}\log B(\alpha+\|F_{ij}\|_1,\beta+K-\|F_{ij}\|_1)\right\}.
\end{aligned}
\end{equation}
Note that we include an extra tuning parameter $\lambda$ in front of the penalty term to 
give us a bit more flexibility in controlling its magnitude.
The function $k\mapsto \log B(\alpha+k,\beta+K-k)$ is convex and takes larger values
when $k$ is close to 0 or $K$ compared to those in between. Using it as a penalty thus favors the set of graphs
which share common edges.

To solve \eqref{eqn:Ftilde-multiple}, we again adopt a minorize-maximization procedure. 
Specifically, write the objective function as
\begin{equation}
f(F^{(1:K)})=\sum_{k=1}^K\sum_{i<j}w_{ij}^{(k)}F_{ij}^{(k)}+\lambda\sum_{i<j}\log B(\alpha+\|F_{ij}\|_1,\beta+K-\|F_{ij}\|_1),
\end{equation}
where $w_{ij}^{(k)}=\hat I (X_i^{(k)};X_j^{(k)})$.
Given a current solution $\check F^{(k)}$, we linearize $f(F^{(1:K)})$ at $\check F^{(k)}$ and get
\begin{align}
f(F^{(1:K)})
\geq
&\sum_{k=1}^K\sum_{i<j}w_{ij}^{(k)}F_{ij}^{(k)}+
\lambda\sum_{i<j}(\|F_{ij}\|-\|\check F_{ij}\|_1)\left(\psi(\alpha+\|\check F_{ij}\|_1)-\psi(\beta+K-\|\check F_{ij}\|_1)\right)\\
=&\sum_{k=1}^K\sum_{i<j}\left(w_{ij}^{(k)}+\lambda\left(\psi(\alpha+\|\check F_{ij}\|_1)-\psi(\beta+K-\|\check F_{ij}\|_1)\right)\right)F_{ij}^{(k)}+C,
\end{align}
where $\psi(x)=\frac{d}{dx}\log\Gamma(x)$ is the digamma function. 
This gives the following weights updating rule:
\begin{equation}
\check w_{ij}^{(k)}=w_{ij}^{(k)}+\lambda\left(\psi(\alpha+\|\check F_{ij}\|_1)-\psi(\beta+K-\|\check F_{ij}\|_1)\right).
\end{equation}
Note that $k\mapsto \psi(\alpha+k)-\psi(\beta+K-k)$ is an increasing function.
Therefore, this updating rule borrows strength across the $K$ graphs---it increases
an edge's weight when $\|F_{ij}\|_1$ is large, i.e., when other graphs also have edge $(i,j)$ present.

\subsection{Algorithms}
As a short conclusion, we summarize the two procedures, which share a lot of similarity but work for different applications,
here in Algorithm \ref{alg:scale-free} and \ref{alg:multiple}. 
After getting the output of the algorithm, we will prune the resulting spanning tree to obtain a forest estimate (to avoid overfitting in high dimensions). 
This can be done by going through the last iteration of the algorithm and stop at the step where the likelihood
is maximized on a held-out dataset. 

\begin{algorithm}[H]
\caption{Scale-free graph estimation}\label{alg:scale-free}
\begin{algorithmic}
\State \textbf{input} Weight matrix $W=(w_{ij})_{d\times d}$, tuning parameter $\lambda$
\State  $F\gets$ output of Algorithm \ref{alg:kruskal} on $W$
\Do
    \State $\check w_{ij}\gets w_{ij}-\frac{\lambda}{\sum_{\ell=1}^d F_{i\ell}}-\frac{\lambda}{\sum_{\ell=1}^d F_{j\ell}}$ 
    \State  $F\gets$ output of Algorithm \ref{alg:kruskal} on $\check W=(\check w_{ij})_{d\times d}$
\doWhile {$F$ has not converged}
\State \textbf{output} $F$
\end{algorithmic}
\end{algorithm}
\vspace{-1.5em}
\begin{algorithm}[H]
\caption{Joint estimation for multiple graphs}\label{alg:multiple}
\begin{algorithmic}
\State \textbf{input} Weight matrices $W^{(k)}=(w_{ij}^{(k)})_{d\times d}$ for $k=1,\dots,K$, tuning parameters $\lambda$, $\alpha$, $\beta$
\State  $F^{(k)}\gets$ output of Algorithm \ref{alg:kruskal} on $(w_{ij}^{(k)})_{d\times d}$ for $k=1,\dots,K$
\Do
    \State $\check w_{ij}^{(k)}\gets w_{ij}^{(k)}+\lambda\left(\psi(\alpha+\| F_{ij}\|_1)-\psi(\beta+K-\| F_{ij}\|_1)\right)$ 
    \State  $F^{(k)}\gets$ output of Algorithm \ref{alg:kruskal} on $\check W^{(k)}=(\check w_{ij})_{d\times d}$ for $k=1,\dots,K$
\doWhile {$F^{(1:K)}$ have not converged}
\State \textbf{output} $F^{(1:K)}$
\end{algorithmic}
\end{algorithm}

\section{Related work}

Before proceeding to present the performance of the proposed nonparametric methods on both simulated and real datasets,
we pause to review some of the existing approaches on estimation of scale-free graphical models and joint estimation of multiple graphical models. 

Most existing methods for estimating graphical models with prior information assume that 
the data follow multivariate Gaussian distributions. 
To encourage a scale-free graph, \citet{liu2011learning} propose to replace the 
$\ell_1$ penalty in the formulation of the graphical lasso by a non-convex power
law regularization term.
Along the same line, \citet{defazio2012convex} impose a convex penalty by using submodular functions and their Lov\'asz extension.
Essentially, both methods try to penalize the log degree of each node, but end up using a continuous/convex surrogate to avoid the combinatorial problems involving the degrees. 
\citet{tan2014learning} propose a general framework to accommodate 
networks with hub nodes, using a convex formulation that involves a row-column
overlap norm penalty.

Methods for inferring Gaussian graphical models on multiple units 
have also been proposed in recent years. 
\citet{Guo11} propose a method for joint estimation of Gaussian graphical models 
by penalizing the graphical lasso objective function by 
the square root of $\ell_1$ norms of the edge vector across all graphs,
which results in a non-convex problem.
A convex joint graphical lasso approach is developed in \cite{Danaher14}, 
which is based on employing generalized fused lasso or group lasso penalties. 
\citet{Peterson15} and \citet{Zhu15} propose Bayesian approaches for inference on multiple Gaussian graphical models. 

We summarize the aforementioned methods, which are to be implemented and compared
in the simulation. Methods proposed in this paper can be viewed as nonparametric counterparts 
to the parametric methods. 

\begin{table*}[h]
\renewcommand\multirowsetup{\centering}
\begin{center}
\begin{tabular}{lccc}
\toprule
\multirow{2}[4]{1cm}{} & \multirow{2}[4]{2cm}{General}  & \multicolumn{2}{c}{With prior information}  \\
      \cmidrule{3-4}
          &             & $\quad$ {Scale-free graph} $\quad$ & $\quad$ {Multiple graphs} $\quad$ \\
      \midrule
\multirow{2}[1]{1cm}{Parametric} & \multirow{2}[1]{2cm}{\texttt{Glasso} \cite{friedman2008sparse}}  & \texttt{SFGlasso}$^*$ \cite{liu2011learning} & \texttt{GuoGlasso}$^*$  \cite{Guo11}\\
 &   & \texttt{HubGlasso}$^\dag$  \cite{tan2014learning} & \texttt{JointGlasso}$^\dag$  \cite{Danaher14} \\
 \midrule
Nonparametric & \texttt{FDE} \cite{liu2011forest} & \texttt{SF-FDE}$^\ddag$ & \texttt{J-FDE}$^\ddag$ \\
\bottomrule
\end{tabular}
\end{center}
\hspace{0.57in}\small{$*$: non-convex method\quad $\dag$: convex method\quad $\ddag$: this paper}
\caption{Summary and comparison between different methods in graphical modeling.}
\label{tab:methods}
\end{table*}

\section{Experiments}

\subsection{Synthetic data}

In this subsection, we evaluate the performance of the proposed methods and other existing methods on synthetic data.

\begin{figure}[!ht]
\begin{center}
\begin{tabular}{cc}
\includegraphics[width=0.3\textwidth]{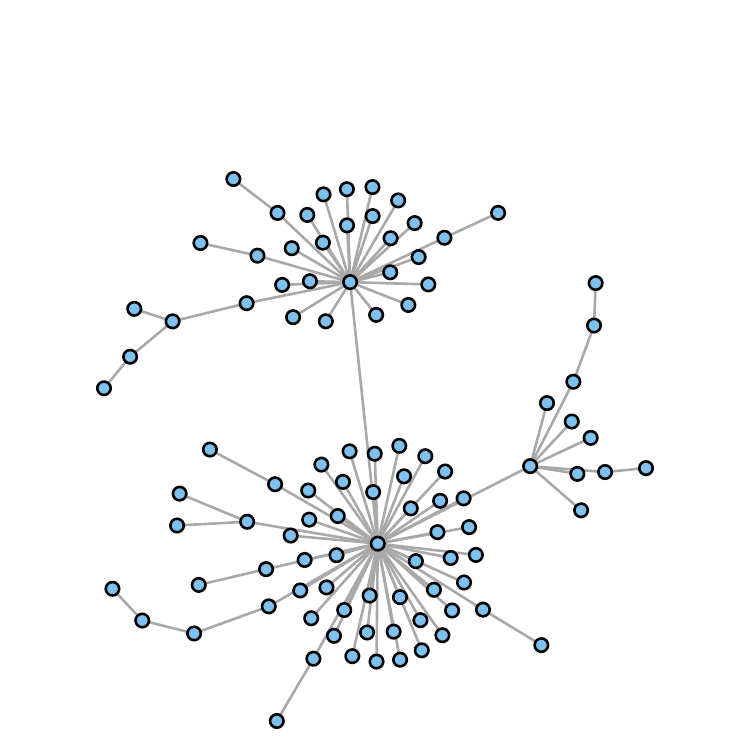} $\quad$ & $\quad$
\includegraphics[width=0.3\textwidth]{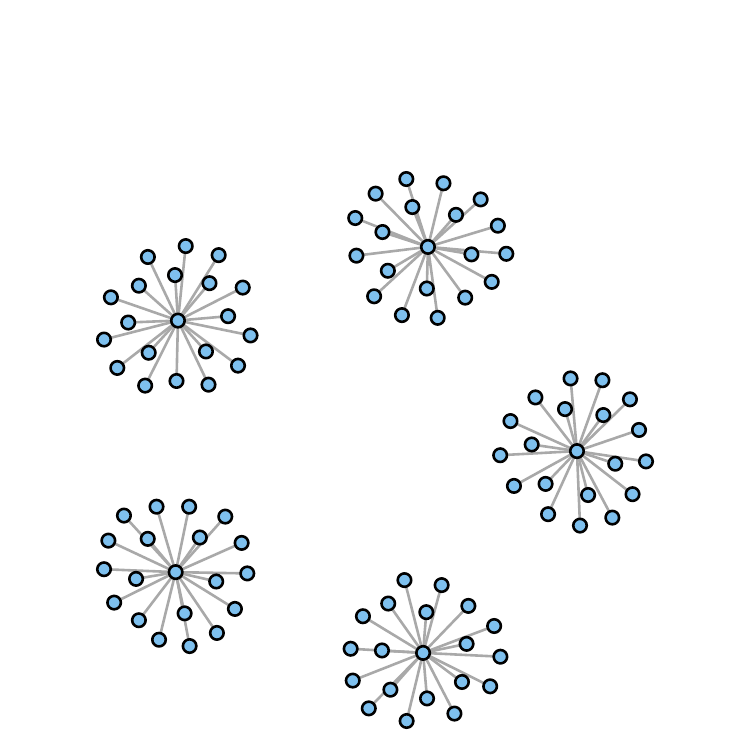}\\
Scale-free graph $\quad$ & $\quad$ Stars
\end{tabular}
\caption{An illustration of simulated graph patterns.}\label{fig:simulation}
\end{center}
\end{figure}

\vspace{-0.5em}
\paragraph{Graph structures} We consider the following types of graph structures with $d=100$ vertices.
\begin{itemize}
\setlength{\parskip}{0pt}
\vspace{-0.5em}
\item \textbf{Scale-free graph}: We use a preferential attachment process to generate a scale-free graph \cite{albert2002statistical}. 
We start with a chain of 4 nodes (i.e., with edges 1--2, 2--3, and 3--4). 
New nodes are added one at a time, and each new node is connected to one existing node with probability $p_i\propto \delta_i^{\alpha}$, where $\delta_i$ is the current degree of the $i$th node, and $\alpha$ is a parameter, which we set to be 1.5 in our experiments. 
A typical realization of such networks is shown in Figure \ref{fig:simulation}~(left).
\item \textbf{Stars}: The graph has 5 stars of size 20; each star is a tree with one root and 19 leaves. 
An illustration is shown in Figure \ref{fig:simulation}~(right).
\item \textbf{Multiple graphs}: We follow the above two mechanisms to generate multiple graphs with similar structures. 
In particular, we generate a set of $K=3$ scale-free graphs, which share 80 common edges
(this is done by applying the above generative model to grow a common tree of size 80 to be shared across the 3 units;
each unit then continues this growing process independently until obtaining a tree of 100 vertices),
and another set of $K=3$ stars graphs, which have 4 common stars and one individual star with distinct roots.
\end{itemize}

\vspace{-1em}

\paragraph{Probability distributions} Given a particular graph, we generate $200$ samples according to two types of probability distributions that are Markov to the graph: Gaussian copulas and $t$ copulas \cite{demarta2005t}.
The Gaussian copula (resp., the $t$ copula) can be thought of as representing the dependence structure implicit in a multivariate Gaussian (multivariate $t$) distribution, while each variable follows a uniform distribution on $[0,1]$ marginally. 
Since the graph structures we consider are trees or forests, we generate the data sequentially, first sampling for an arbitrary node in a tree, and then drawing samples for the neighboring nodes according to the conditional distribution given by the copula until going through all nodes in the tree. 
In our simulations, the degree of freedom of the $t$ copula is set to be 1, and the correlation coefficients are chosen to be 0.4 and 0.25 for the Gaussian and the $t$ copula. 

\vspace{-0.5em}
\paragraph{Methods} 
We implement methods that are summarized in Table \ref{tab:methods}.
For the forest-based methods, we use a held-out set of size $100$ to select tuning parameter and prune the estimated spanning trees.
To implement the Gaussian-based methods, we first transform the data marginally to be approximately Gaussian.
We choose the tuning parameters by searching through a fine grid, and selecting those that maximize the likelihood on the held-out set. We refer to this as \emph{held-out tuning}. 
The results obtained by the held-out tuning reflect the performance of the methods in a fully data-driven way.
In addition, we also consider what we call \emph{oracle tuning}, where the tuning parameters are chosen to maximize the $F_1$ score of the estimated graph. 
This tuning method requires the knowledge of the true graph, and hence it's not obvious that there would exist a data-driven way to achieve this.
We include the oracle tuning as a way to show the optimal performance possibly achieved by the methods. 

\vspace{-0.5em}\paragraph{Results} 
For both scale-free graphs and multiple graphs, 
we carry out four sets of experiments, with data generated from the two types of graphs and the two types of distributions. 
For each set of experiments, we repeat the simulations 10 times and record the $F_1$ scores of the estimated graphs for each method. 
An $F_1$ score is the harmonic mean of a method's precision and recall and hence a measure of its accuracy. 
It's a number between 0 and 1; a higher score means better accuracy and 1 means perfect labelling. 
The average $F_1$ scores are shown in Table \ref{tab:f1score}.
From the table, we see that \texttt{SF-FDE} and \texttt{J-FDE} always outperform \texttt{FDE} on these particular situations.
Also, \texttt{SF-FDE} and \texttt{FDE} perform better than the other three methods using held-out tuning as the penalized likelihood methods tend to select more edges when tuning parameters are chosen to maximize the held-out likelihood. 
When the true copula is Gaussian, the graphical-lasso-based methods all have very high scores if oracle tuning is used; they fail to deliver good performance when the true copula is no longer Gaussian. 
On the other hand, the FDE-based methods are not affected too much by the true distribution.

\begin{table*}[t]
\renewcommand\multirowsetup{\centering}
\centering
\begin{tabular}{lcccccccc}
\textbf{Graphs with hubs} \\
\toprule
\multirow{2}[4]{2cm}{Graph $\times$ Dist.} & \multirow{2}[4]{1.2cm}{\texttt{FDE}} & \multirow{2}[4]{1.2cm}{\texttt{SF-FDE}} & \multicolumn{2}{c}{\texttt{Glasso}} & \multicolumn{2}{c}{\texttt{SFGlasso}} &  \multicolumn{2}{c}{\texttt{HubGlasso}} \\
      \cmidrule{4-9}
          &       &       & \scriptsize{held-out} & \scriptsize{oracle} & \scriptsize{held-out} & \scriptsize{oracle} & \scriptsize{held-out} & \scriptsize{oracle}\\
      \midrule
Scale-free $\times\ \mathcal N$  & 0.79 & 0.92 & 0.24 & 0.91 & 0.42 & 0.92 & 0.16 & 0.88 \\
Stars $\times\ \mathcal N$ & 0.82 & 0.96 & 0.25 & 0.93 & 0.46 & 0.98 & 0.08 & 0.99 \\
Scale-free $\times\ t$ & 0.89 & 0.98 & 0.30 & 0.43 & 0.47 & 0.53 & 0.07 & 0.55 \\
Stars $\times\ t$ & 0.93 & 0.98 & 0.32 & 0.56 & 0.50 & 0.67 & 0.09 & 0.79\\
\bottomrule
\\
\textbf{Multiple graphs} \\
\toprule
\multirow{2}[4]{2cm}{Graph $\times$ Dist.} & \multirow{2}[4]{1.2cm}{\texttt{FDE}} & \multirow{2}[4]{1.2cm}{\texttt{J-FDE}} & \multicolumn{2}{c}{\texttt{Glasso}} &  \multicolumn{2}{c}{\texttt{GuoGlasso}} & \multicolumn{2}{c}{\texttt{JointGlasso}} \\
      \cmidrule{4-9}
          &       &       & \scriptsize{held-out} & \scriptsize{oracle} & \scriptsize{held-out} & \scriptsize{oracle} & \scriptsize{held-out} & \scriptsize{oracle}\\
      \midrule
Scale-free $\times\ \mathcal N$ & 0.78 & 0.90 & 0.25 & 0.92 & 0.84 & 0.97 & 0.17 & 0.97 \\
Stars $\times\ \mathcal N$ & 0.80 & 0.92 & 0.26 & 0.92 & 0.80 & 0.95 & 0.17 & 0.96 \\
Scale-free $\times\ t$ & 0.91 & 0.98 & 0.30 & 0.44 & 0.61 & 0.64 & 0.23 & 0.66 \\
Stars $\times\ t$ & 0.92 & 0.98 & 0.33 & 0.53  & 0.66 & 0.70 & 0.27 & 0.71\\
\bottomrule
\end{tabular}
\caption{Averaged $F_1$ scores for methods applied on the simulated data.}\label{tab:f1score}
\end{table*}

\subsection{Real data}

\vspace{-0.5em}\paragraph{Stock price data} We test our methods on
the daily closing prices for $d=417$ stocks that are constantly in the S\&P 500 index from Yahoo!~Finance.
The log returns of each stock are replaced by their respective normal scores, subject to a Winsorized truncation.

\begin{figure}[!ht]
\begin{center}
\begin{tabular}{cc}
\includegraphics[width=0.47\textwidth]{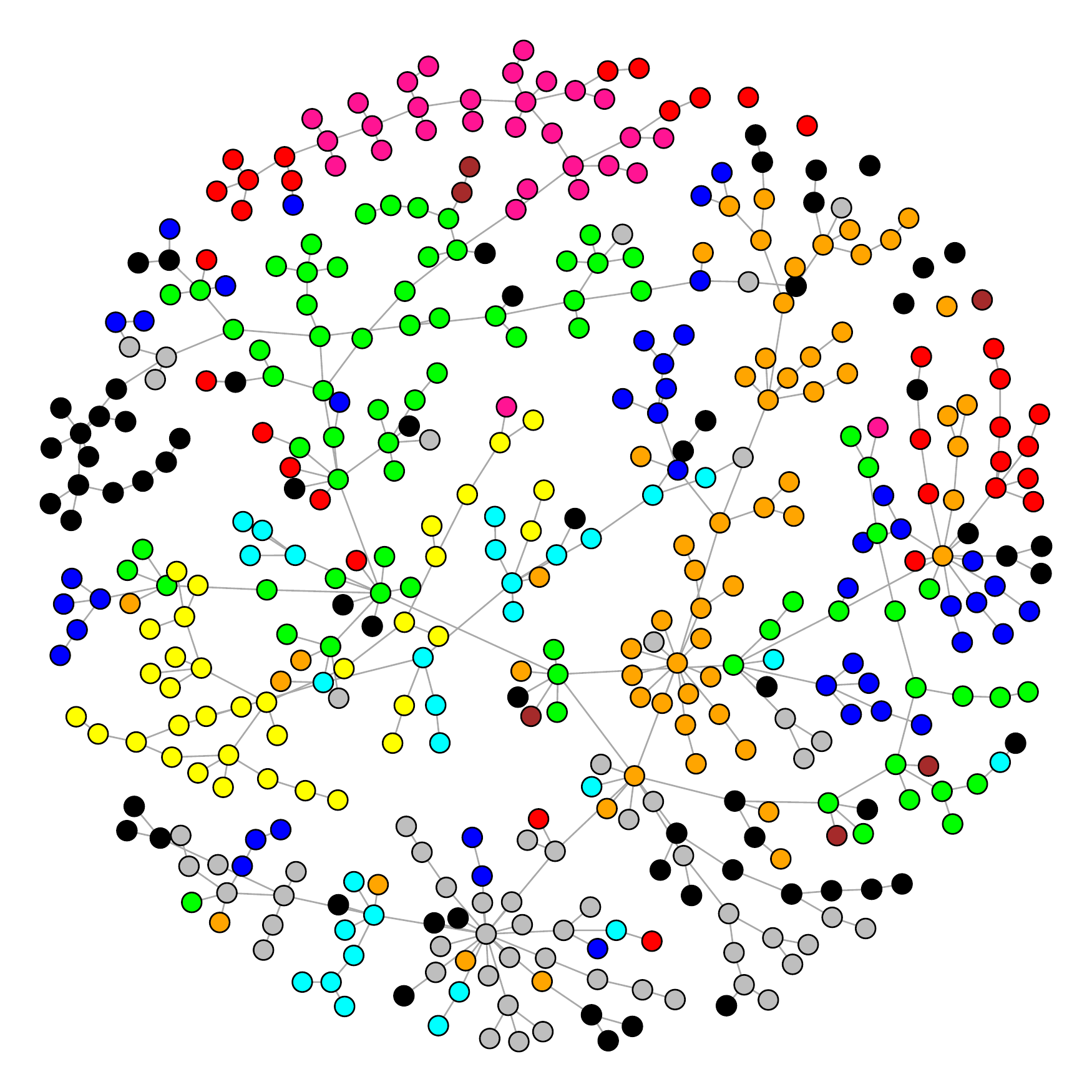} $\quad$ &
\includegraphics[width=0.47\textwidth]{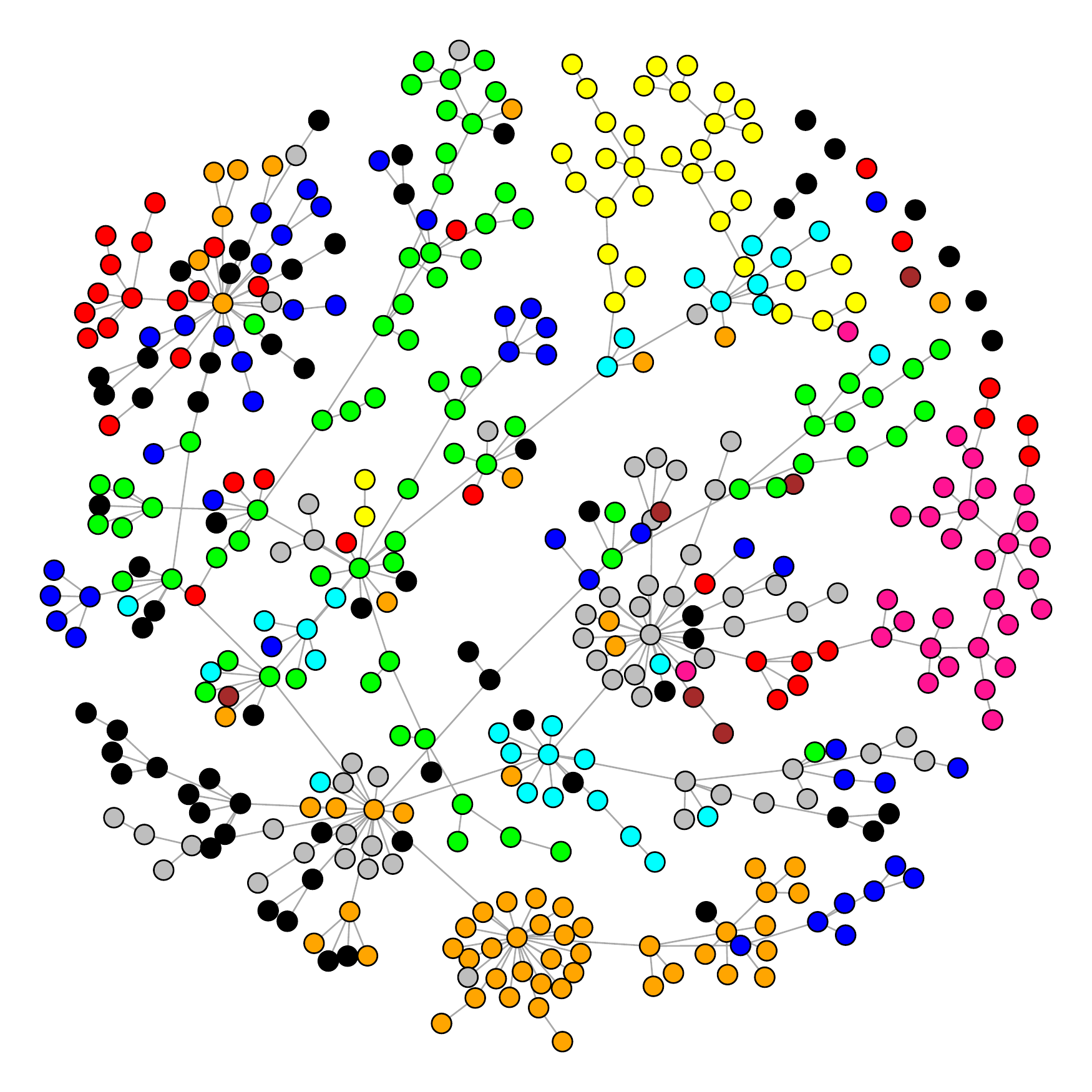}\\
\texttt{FDE} $\quad$ & \texttt{SF-FDE}
\end{tabular}
\caption{Estimated graphs for \texttt{FDE} and \texttt{SF-FDE} applied on the stock price data. The stocks are colored according to their Global Industry Classification Standard categories.}\label{fig:stock}
\end{center}
\end{figure}

In the application of learning scale-free forests, we use the data from the first 9 months of 2014 as the training data 
and the data from the last 3 months of 2014 as the held-out data.
The result turns out that \texttt{SF-FDE} yields a larger held-out log-likelihood than \texttt{FDE} (64.5 compared to 62.6),
implying that a scale-free approximation is helpful in predicting the relationships. 
The estimated graphs by \texttt{FDE} and \texttt{SF-FDE} are shown in Figure~\ref{fig:stock}.
We see that the resulting clusters by \texttt{SF-FDE} tend to be more consistent with the Global Industry Classification Standard categories,
which are indicated by different colors in the graph.

\begin{figure}[ht]
\begin{center}
\vspace{-0.5cm}
\begin{tabular}{cccc}
\includegraphics[width=1\textwidth]{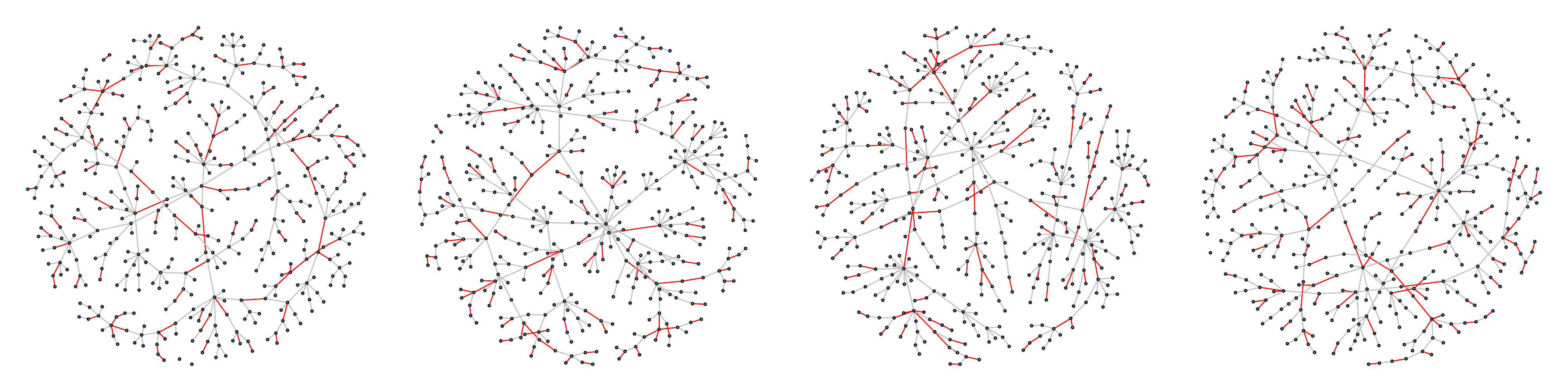} \\
(a) 2009 \qquad\qquad\qquad\; (b) 2010 \qquad\qquad\qquad\;\; (c) 2011 \qquad\qquad\qquad\; (d) 2012 \\
\end{tabular}
\end{center}
\caption{Estimated graphs for $\texttt{J-FDE}$ applied on the stock price data. Common edges across the 4 graphs are colored in red.}
\label{fig:stock_multiple}
\end{figure}

We also consider the application of learning multiple forests by dividing the data into 4 periods from 2009 to 2012, one for a year, and model the 4-unit data using our proposed method. The aggregated held-out log-likelihood over the 4 units are 193.4 for \texttt{J-FDE} and 185.5 for \texttt{FDE}. The numbers of common edges across the 4 graphs are 111 for \texttt{J-FDE} and 24 for \texttt{FDE}, respectively. Figure~\ref{fig:stock_multiple} shows the estimated graphs by \texttt{J-FDE}, where common edges across the 4 units are colored in red.

\paragraph{University webpage data} As a second example, we apply our methods to the university webpage data from the ``World Wide Knowledge Base''
project at Carnegie Mellon University, which consists of the occurrences of various terms on student webpages from
4 computer science departments at Texas, Cornell, Washington, and Wisconsin. We choose a subset of 100 terms with the largest entropy. In the analysis, we compute the empirical distributions instead of kernel density estimates since the data is discrete.

\begin{figure}[!ht]
\begin{center}
\begin{tabular}{cc}
\includegraphics[width=0.46\textwidth]{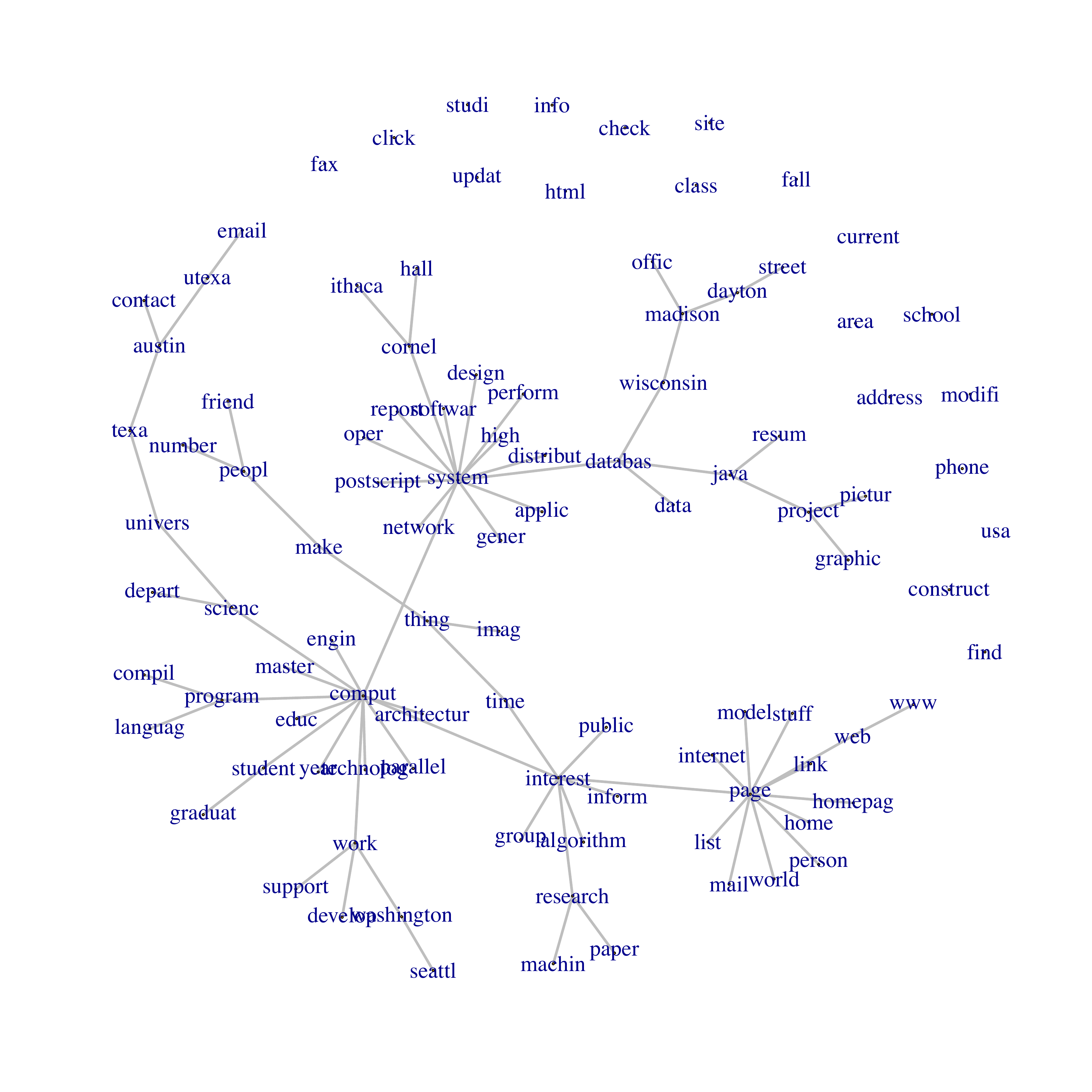} $\quad$ &
\includegraphics[width=0.46\textwidth]{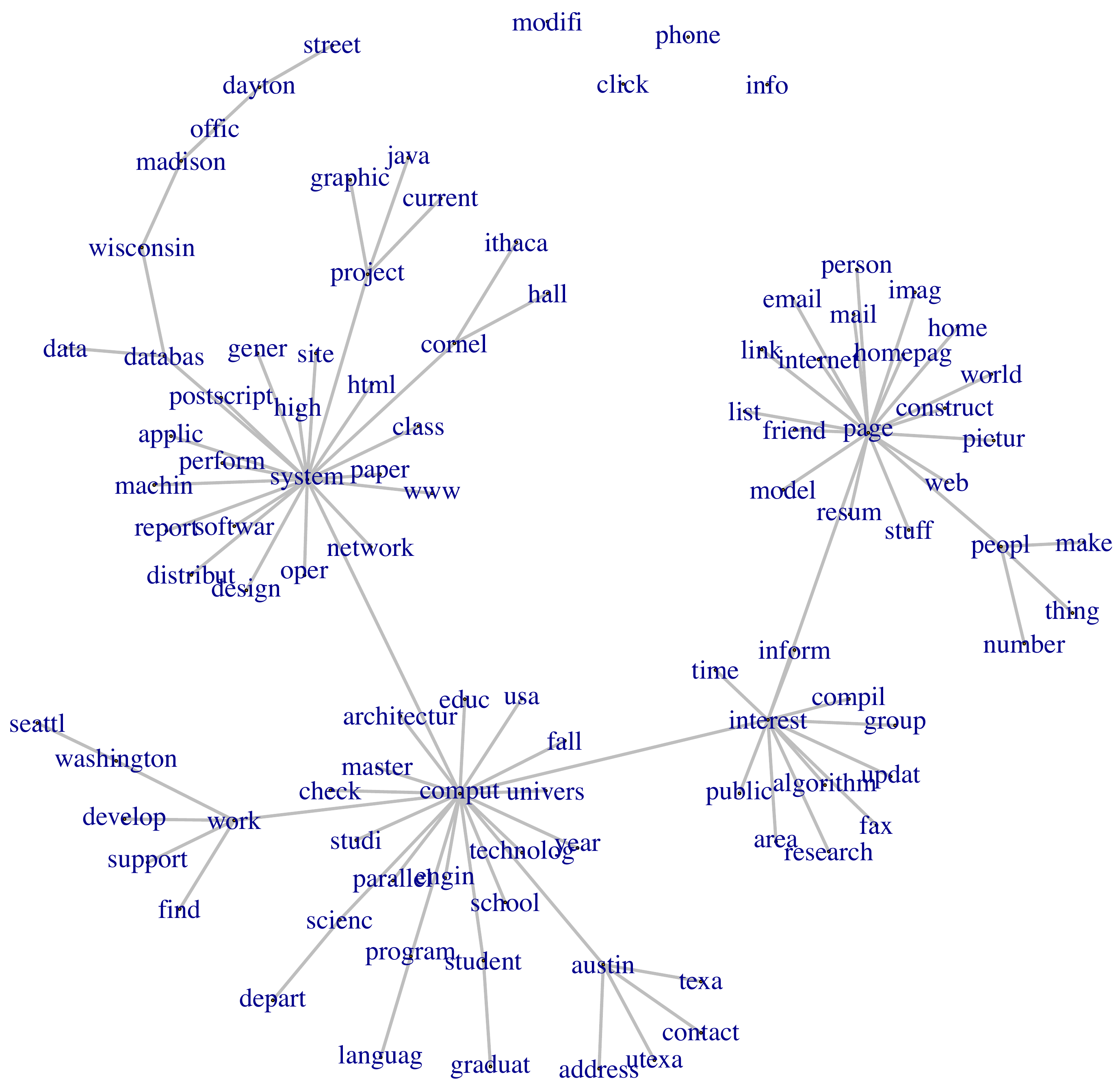}\\
\texttt{FDE} $\quad$ & \texttt{SF-FDE}
\end{tabular}
\caption{Estimated graphs for \texttt{FDE} and \texttt{SF-FDE} applied on the university webpage data.}\label{fig:web}
\end{center}
\end{figure}

\begin{figure}[H]
\begin{center}
\begin{tabular}{cccc}
\includegraphics[width=0.216\textwidth]{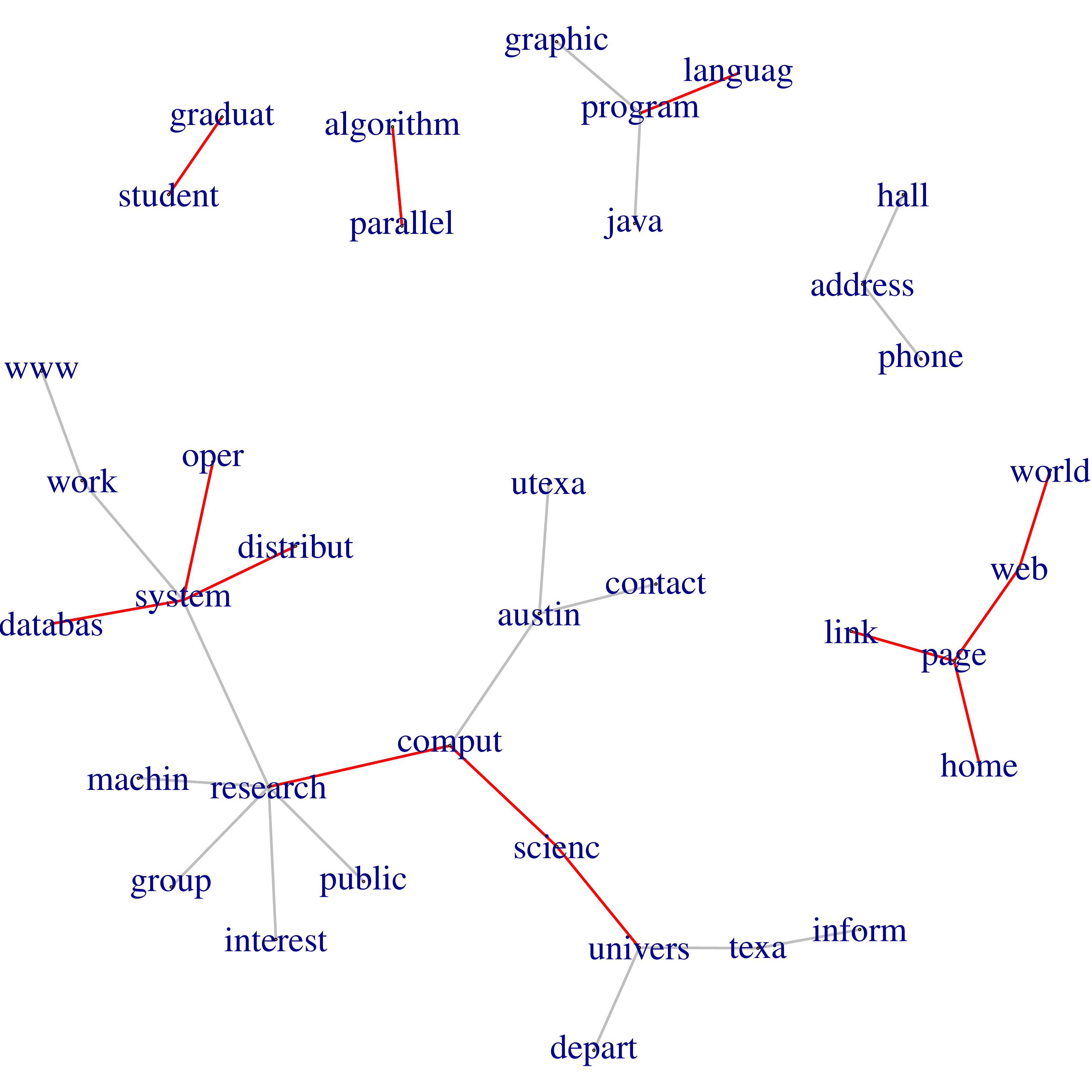} \; &
\includegraphics[width=0.216\textwidth]{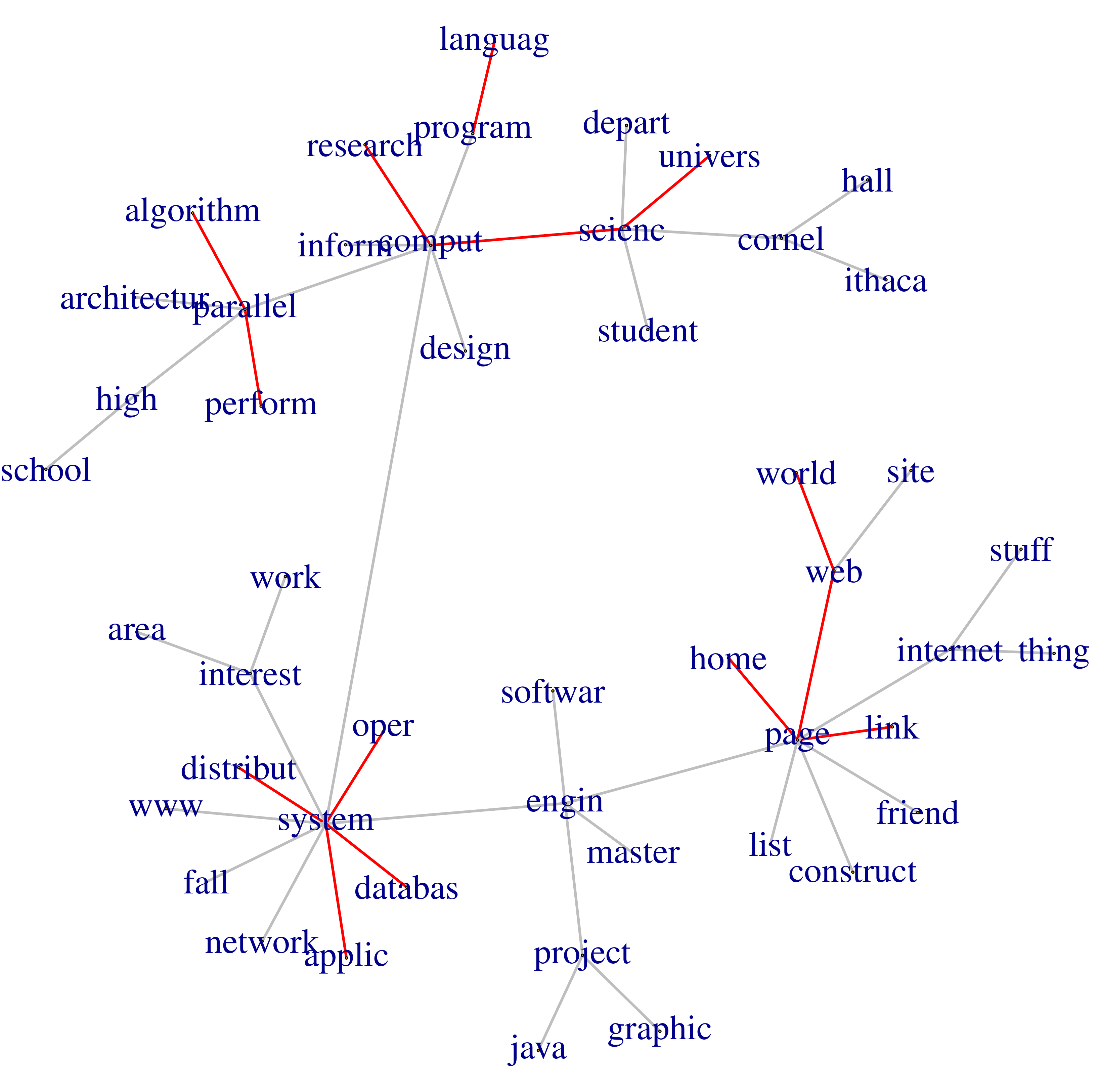} \; &
\includegraphics[width=0.216\textwidth]{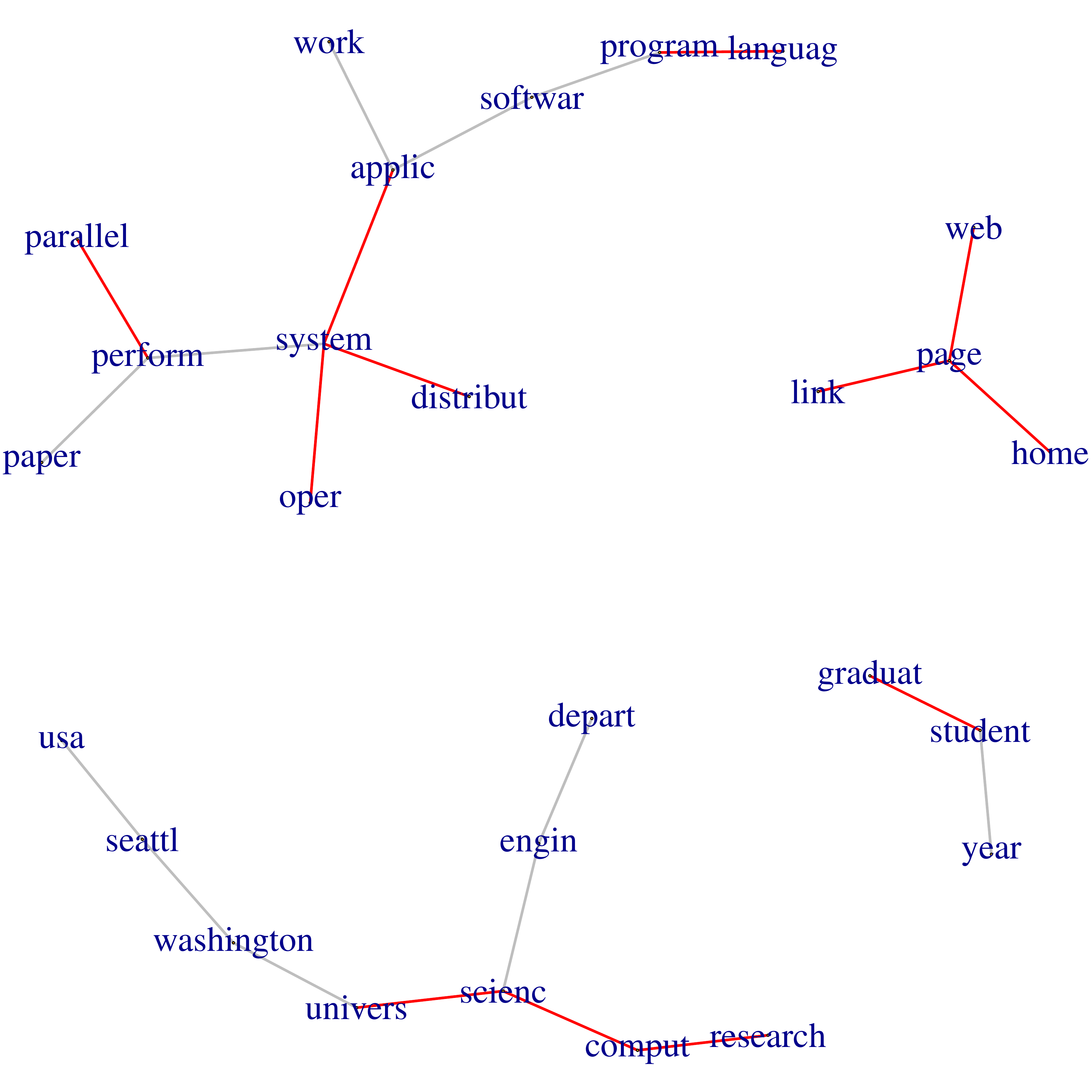} \;&
\includegraphics[width=0.216\textwidth]{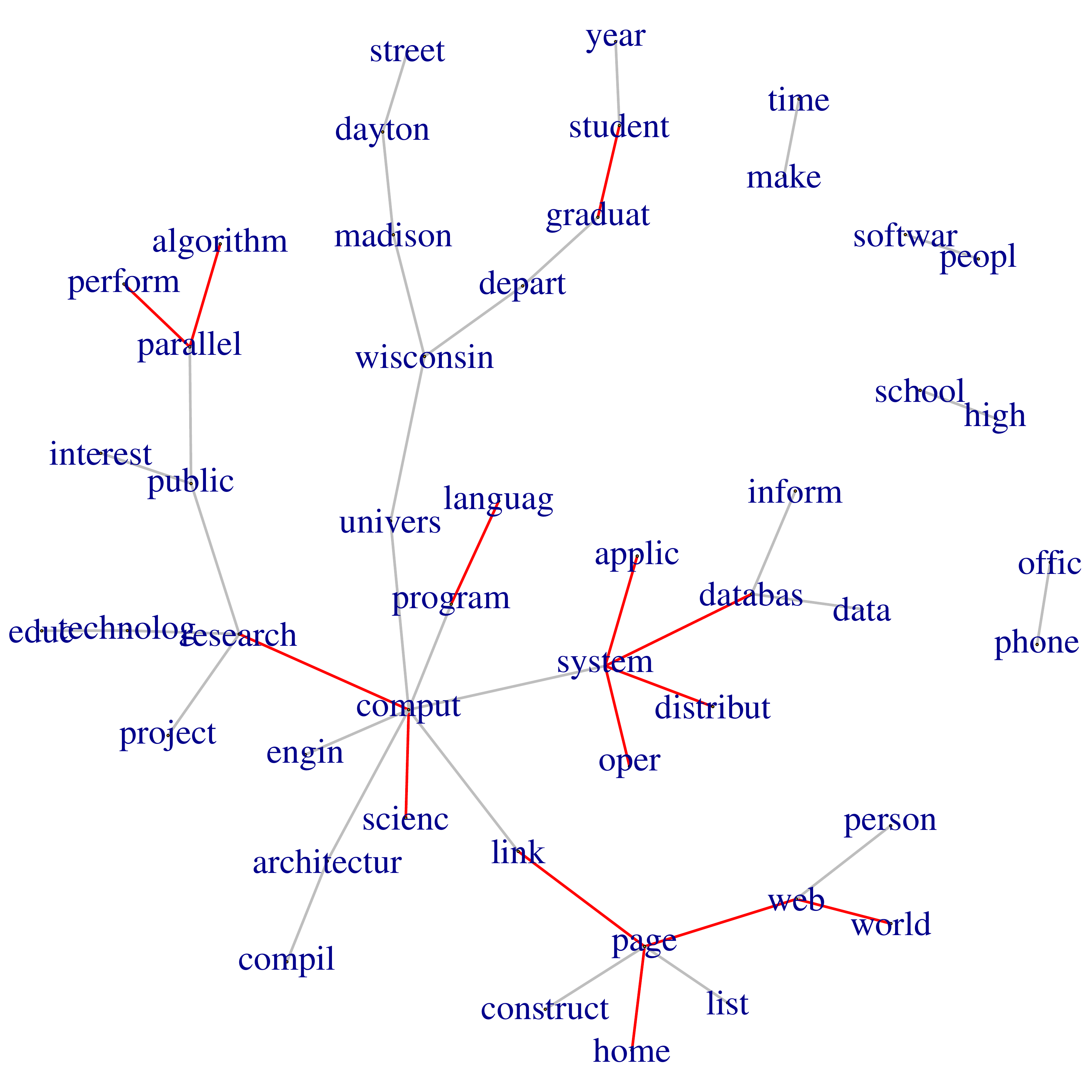}\\
(a) Texas \; & (b) Cornel \; & (c) Washington \; & (d) Wisconsin \\
\end{tabular}
\caption{Estimated graphs for \texttt{J-FDE} applied on the university webpage data. Edges shared by at least 3 units are colored in red.}\label{fig:web_multiple}
\end{center}
\end{figure}

To understand the relationships among the terms, we first wish to identify terms that are hubs. Figure~\ref{fig:web} shows that \texttt{SF-FDE} detects 4 highly connected nodes of degree greater than 10: \emph{comput}, \emph{system}, \emph{page}, and \emph{interest}. Then we model the 4-unit data, one for a university. 
Figure~\ref{fig:web_multiple} shows the estimated graphs by \texttt{J-FDE} (isolated nodes are not displayed in each graph). These results provides an intuitive explanation of the relationships among the terms across the 4 universities.

\section{Conclusion}
In this paper, we introduce a nonparametric framework for incorporating prior knowledge to assist estimation of graphical models.
Instead of Gaussianity assumptions, it assumes the density is Markov to a forest, thus allowing arbitrary distribution. 
A key ingredient is to design a prior distribution on graphs that favors those consistent with the prior belief.
We illustrate the idea by proposing such prior distributions, which lead to two algorithms, 
for the problems of estimating scale-free networks and multiple graphs with similar structures.
An interesting future direction is to apply this idea to more applications and different types of prior information. 

\bibliographystyle{plainnat}
\bibliography{forestPrior}

\begin{thebibliography}{15}
\providecommand{\natexlab}[1]{#1}
\providecommand{\url}[1]{\texttt{#1}}
\expandafter\ifx\csname urlstyle\endcsname\relax
  \providecommand{\doi}[1]{doi: #1}\else
  \providecommand{\doi}{doi: \begingroup \urlstyle{rm}\Url}\fi

\bibitem[Albert and Barab{\'a}si(2002)]{albert2002statistical}
R{\'e}ka Albert and Albert-L{\'a}szl{\'o} Barab{\'a}si.
\newblock Statistical mechanics of complex networks.
\newblock \emph{Reviews of Modern Physics}, 74\penalty0 (1):\penalty0 47, 2002.

\bibitem[Chow and Liu(1968)]{chow1968approximating}
C~Chow and C~Liu.
\newblock Approximating discrete probability distributions with dependence
  trees.
\newblock \emph{Information Theory, IEEE Transactions on}, 14\penalty0
  (3):\penalty0 462--467, 1968.

\bibitem[Danaher et~al.(2014)Danaher, Wang, and Witten]{Danaher14}
Patrick Danaher, Pei Wang, and Daniela~M Witten.
\newblock The joint graphical lasso for inverse covariance estimation across
  multiple classes.
\newblock \emph{Journal of the Royal Statistical Society: Series B (Statistical
  Methodology)}, 76\penalty0 (2):\penalty0 373--397, 2014.

\bibitem[Defazio and Caetano(2012)]{defazio2012convex}
Aaron Defazio and Tiberio~S Caetano.
\newblock A convex formulation for learning scale-free networks via submodular
  relaxation.
\newblock In \emph{Advances in Neural Information Processing Systems}, pages
  1250--1258, 2012.

\bibitem[Demarta and McNeil(2005)]{demarta2005t}
Stefano Demarta and Alexander~J McNeil.
\newblock The \emph{t} copula and related copulas.
\newblock \emph{International Statistical Review}, 73\penalty0 (1):\penalty0
  111--129, 2005.

\bibitem[Friedman et~al.(2008)Friedman, Hastie, and
  Tibshirani]{friedman2008sparse}
Jerome Friedman, Trevor Hastie, and Robert Tibshirani.
\newblock Sparse inverse covariance estimation with the graphical lasso.
\newblock \emph{Biostatistics}, 9\penalty0 (3):\penalty0 432--441, 2008.

\bibitem[Guo et~al.(2011)Guo, Levina, Michailidis, and Zhu]{Guo11}
Jian Guo, Elizaveta Levina, George Michailidis, and Ji~Zhu.
\newblock Joint estimation of multiple graphical models.
\newblock \emph{Biometrika}, 98\penalty0 (1):\penalty0 1--15, 2011.

\bibitem[Hunter and Lange(2004)]{hunter2004tutorial}
David~R Hunter and Kenneth Lange.
\newblock A tutorial on {MM} algorithms.
\newblock \emph{The American Statistician}, 58\penalty0 (1):\penalty0 30--37,
  2004.

\bibitem[Kruskal(1956)]{kruskal1956shortest}
Joseph~B Kruskal.
\newblock On the shortest spanning subtree of a graph and the traveling
  salesman problem.
\newblock \emph{Proceedings of the American Mathematical society}, 7\penalty0
  (1):\penalty0 48--50, 1956.

\bibitem[Liu et~al.(2011)Liu, Xu, Gu, Gupta, Lafferty, and
  Wasserman]{liu2011forest}
Han Liu, Min Xu, Haijie Gu, Anupam Gupta, John Lafferty, and Larry Wasserman.
\newblock Forest density estimation.
\newblock \emph{The Journal of Machine Learning Research}, 12:\penalty0
  907--951, 2011.

\bibitem[Liu and Ihler(2011)]{liu2011learning}
Qiang Liu and Alexander~T Ihler.
\newblock Learning scale free networks by reweighted $\ell_1$ regularization.
\newblock In \emph{International Conference on Artificial Intelligence and
  Statistics}, pages 40--48, 2011.

\bibitem[Peterson et~al.(2015)Peterson, Stingo, and Vannucci]{Peterson15}
Christine Peterson, Francesco~C. Stingo, and Marina Vannucci.
\newblock Bayesian inference of multiple {G}aussian graphical models.
\newblock \emph{Journal of the American Statistical Association}, 110\penalty0
  (509):\penalty0 159--174, 2015.

\bibitem[Tan et~al.(2014)Tan, London, Mohan, Lee, Fazel, and
  Witten]{tan2014learning}
Kean~Ming Tan, Palma London, Karthik Mohan, Su-In Lee, Maryam Fazel, and
  Daniela Witten.
\newblock Learning graphical models with hubs.
\newblock \emph{The Journal of Machine Learning Research}, 15\penalty0
  (1):\penalty0 3297--3331, 2014.

\bibitem[Tang et~al.(2015)Tang, Sun, and Xu]{tang2015learning}
Qingming Tang, Siqi Sun, and Jinbo Xu.
\newblock Learning scale-free networks by dynamic node specific degree prior.
\newblock In \emph{Proceedings of the 32nd International Conference on Machine
  Learning}, pages 2247--2255, 2015.

\bibitem[Zhu and Barber(2015)]{Zhu15}
Yuancheng Zhu and Rina~Foygel Barber.
\newblock The log-shift penalty for adaptive estimation of multiple {G}aussian
  graphical models.
\newblock \emph{\emph{In} Proceedings of the 18th International Conference on
  Artificial Intelligence and Statistics}, pages 1153--1161, 2015.

\end{thebibliography}

\end{document}